\documentclass[12pt]{iopart}

\usepackage{epsfig}
\usepackage{iopams}
\newcommand{\JSTAT}{Journal of Statistical Mechanics -Theory and Experiments}

\begin{document}

\title[RANDOM WALKS ESTIMATE LAND VALUE]{RANDOM WALKS ESTIMATE LAND VALUE IN CITIES}

\author{Ph. Blanchard \& D. Volchenkov}

\address{Bielefeld -- Bonn Stochastic Center, University of Bielefeld, Postfach 100131,D-33501, Bielefeld, Germany}
\ead{blanchard@physik.uni-bielefeld.de,  volchenk@physik.uni-bielefeld.de}

\begin{abstract}

 Expected urban population doubling calls for a compelling theory of the city.
  Random walks and diffusions defined on spatial city graphs spot hidden areas
  of geographical isolation in the urban landscape going downhill.
 First--passage time to a place correlates with assessed value of land in that.
The
method accounting the
    average number of random turns at junctions
on the way to reach any
     particular place in the city from various starting points could be used
     to identify isolated neighborhoods in big cities with a complex web of roads,
     walkways and public transport systems.

\end{abstract}


\submitto{\JSTAT}


\section{The problem of massive urbanization}
\label{sec:massive}
\noindent

Multiple increases in urban population
that had occurred in Europe
at the beginning of the $20^{\mathrm{th}}$ century
made urban agglomerations suffer
from the problems of urban decay
such as wide-spread poverty,
high unemployment,
and rapid changes
in the racial composition of neighborhoods.
Riots and social revolutions
happened in response
to conditions of urban decay
in many European countries established
regimes affecting immigrants
and certain population groups
{\it de facto} alleviating
the burden of the haphazard urbanization
by increasing its deadly price.

More than half of world's population, 3.3 billion people, is now
 living in cities, and the figure is about to double by 2030 \cite{UN:2007}.
Urban sprawl in US covered $41,000$ $\mathrm{km}^{2}$ in 20 years,
the area that equals that of the state of Switzerland \cite{USsprawl}.
In Europe, urban sprawl has covered $8,000$ $\mathrm{km}^{2}$
 in 10 years the area that equals the territory of Luxembourg \cite{EEA:2006}.
Unsustainable pressure on resources causes the increasing loss
 of fertile lands through degradation and climate changes
 through the increasing carbon emissions warming the earth's atmosphere.
City development planners will face great challenges in preventing
cities from unlimited expansion.

Global poverty is in flight becoming
a primarily urban phenomenon in the developing world.
2 billion new urban settlers in the next
20 years will live in slums on no more than \$1 a day,
 adding to 1 billion already living there \cite{Ravallion:2007}.
Faults in urban planning, poverty, redlining,
immigration restrictions and clustering of
minorities dispersed over the spatially isolated
pockets of streets trigger urban decay,
a process by which a city falls into a state of disrepair.
The speed and scale of urban growth in
that require urgent global actions to help
 cities prepare for growth and to avoid them of
being the future epicenters of poverty and human suffering.

\section{The scope of study and conclusions}

In the present paper, we investigate
spatial graphs of
several compact urban patterns: the city canal networks
of Venice and Amsterdam, as well as
the almost regular array of streets in Manhattan.

In the next section (Sec.~\ref{sec:Networks}), we briefly discuss
the principles which is are used for encoding urban environments
into geometrical  graphs. In Sec.~\ref{sec:Theory}, we develop the
theory of linear automorphisms for undirected graphs and
demonstrate that they can be interpreted as random walks on
graphs. By the way, many geometrical properties of graph may
acquire probabilistic interpretations. In particular, we are
interested in first--passage times to the nodes of graphs -- the
expected number of steps required for random walkers starting from
an arbitrary node of the graph to reach the given node for the
first time. The first-passage time is the important characteristic
of the node quantifying its isolation in the complex graph
structure.

Sociologists think that isolation worsens an area's economic
prospects by reducing opportunities for commerce, and engenders
 a sense of isolation in inhabitants, both of which can fuel
poverty and crime. It is well known that many social variables
demonstrate striking spatial distribution patterns, and therefore
may be detected and predicted by a structural analysis. In the
last   Sec.~\ref{sec:urban}, we discuss two examples of such a
type: the spatial  isolation of Venetian Ghetto, the small area of
Venice in which Jewish people were compelled to live under the
Venetian Republic, and the value of land in the urban pattern of
Manhattan. In particularly, we compare the tax assessment rates of
places in Manhattan with first-passage times to them. The data
positively relate the geographic accessibility of places in
Manhattan
 with their 'unearned increments' quantifying the level
of appreciation in value of the site.

The proposed method  could be used
     to identify isolated neighborhoods in big cities with a complex web of roads,
     walkways and public transport systems.

\section{Spatial networks of urban environments}
\label{sec:Networks}
\noindent

In traditional
 urban researches, the dynamics of an urban pattern come
 from the landmasses, the physical aggregates of buildings
 delivering place for people and their activity.
The relationships between certain components of the urban texture
  are often measured along streets and routes considered as edges
  of a planar graph, while the traffic end points and street
  junctions are treated as nodes. Such a   primary graph
  representation of urban networks is grounded on relations
  between junctions through the segments of streets. The usual
   city map based on Euclidean geometry can be considered as an
   example of primary city graphs.

In space syntax theory (see \cite{Hillier:1984,Hillier:1999}),
 built environments are treated as systems
of spaces of vision subjected to a configuration analysis.
Being
irrelevant to the physical distances, spatial graphs
 representing the
urban environments are
removed from the physical space.
It has been demonstrated in multiple experiments
that spatial perception
shapes peoples understanding of how
a place is organized and eventually  determines the pattern of local
 movement, \cite{Hillier:1999}.
The aim
of the space syntax  study is to estimate the relative proximity
between different locations and to associate these distances to
 the densities of
human activity along the links connecting them,
\cite{Hansen:1959,Wilson:1970,Batty:2004}. The surprising accuracy
of predictions of human behavior in cities based on the purely
topological analysis of different urban street layouts within the
space syntax approach attracts meticulous attention
\cite{Penn:2001}.

The decomposition of
urban spatial networks
into the complete sets
of intersecting open spaces
can be based on a number of different principles.
In  \cite{Jiang:2004},
while identifying a street over a plurality of routes
on a city map, the  named-street approach has been used, in
which two different arcs of the primary city network were
assigned to the same identification number (ID) provided they share the same
street name.

In our paper, we take a "named-streets"-oriented point of view
on the decomposition of
urban spatial networks
into the complete sets
of intersecting open spaces
 following our previous works \cite{Volchenkov:2007a,Volchenkov:2007b}.
 Being interested in the statistics of random walks defined on spatial
networks of urban patterns, we assign an individual
street ID code to each continuous segment of a street. The spatial
 graph of urban environment is then constructed by
mapping all edges (segments of streets) of the city map
shared the same street ID into nodes
 and all intersections among each pair of edges of the primary graph
into the edges of the secondary graph connecting the corresponding nodes.

\section{Theory of linear automorphisms for undirected graphs}
\label{sec:Theory}
\noindent

\subsection{Graphs and their linear automorphisms}
\label{subsec:graphs}
\noindent

Consider a transitive permutation group $\mathcal{P}(V) $
defined on a finite ordered set of objects $V$, $|V|=N$.
Its representation
consists of all
  $N\times N$ orthogonal matrices
${\bf \Pi}_{\sigma}$ such that
  $\left({\bf \Pi}_{\sigma}\right)_{i,\,\sigma i}=1$,
and
 $\left({\bf \Pi}_{\sigma}\right)_{i,j}=0$
if $j\ne \sigma i$.
We can define an induced action of $\mathcal{P}(G)$
on the set of all
2-subsets $V\times V$
by
\begin{equation}
\label{trnasitive}
\sigma(v,w)\,=\, \left(\sigma v,\sigma w\right),\quad \sigma\in \mathcal{P}(V),
\quad v,w \in V.
\end{equation}
Given a binary relation of adjacency "$\sim$" defined on the
finite set $V$, we denote its graph by $G$ and
conveniently represent by its adjacency matrix $\bf A$ --
the $N\times N$ matrix
where the non-diagonal entry $A_{ij}=1$ if and only if $i\sim j$,
and is zero otherwise.

Among all possible permutation
groups $\mathcal{P}(V)$ defined over the set $V$,
there is one 'compatible' with the binary relation
 "$\sim$" that is the
 automorphism group of the graph $\mathrm{Aut}(G)$
  such that
for any
 $\sigma\in \mathrm{Aut}(G)$,
$u\sigma\sim v\sigma$, if and only if $u\sim v$.
A permutation $\sigma \in \mathrm{Sym}(G)$
 belongs to $\mathrm{Aut}(G)$ if and only if
its matrix representation $\Pi_\sigma$
commutes with the adjacency matrix
of $G$ \cite{Chan:1997},
\begin{equation}
\label{chan}
\left[{\bf A},{\bf \Pi}_\sigma\right]\,=\,0.
\end{equation}
A linear function of the graph $G$ defined on its adjacency matrix,
\begin{equation}
\label{lin_fun}
f\left({\bf A}\right)_{ij}\,=
\,\sum_{\,s,l=1}^N\, F_{ij,sl}\,A_{sl}, \quad F_{ij,sl}\,\in \,\mathbb{R},
\end{equation}
belongs to the representation of
 the automorphism group
$\mathrm{Aut}(G)$
in the group of linear transformations (\ref{lin_fun})
if
\begin{equation}
\label{permut_invar}
{\bf \Pi}_{\sigma}^\top\, f\left({\bf A}\right)\,{\bf \Pi}_{\sigma}\,=\,
f\left({\bf \Pi}_{\sigma}^\top\,{\bf A}\,{\bf \Pi}_{\sigma}\right)
\quad \& \quad
\sum_{j\in V}\,A_{ij}\, \,=\,\sum_{j\in V}\,f\left({\bf A}\right)_{ij},
\end{equation}
for any $\sigma\in\mathrm{Aut}(G)$. In \cite{Volchenkov:2008}, we
have shown that any function defined on a simple undirected graph
$G$ and satisfying (\ref{permut_invar}) can be associated to
  a stochastic process,
\begin{equation}
\label{T_beta}
f\left(\bf{A}\right)_{ij}\,=\,
(1-\beta)\,\delta_{ij}\, +
\, \frac{\,\,\beta\,\,}{\,\,\kappa_i\,\,}A_{ij}, \quad
\kappa_i\,\equiv\,\sum_{j\in V}\,A_{ij},
\end{equation}
where  $\beta\in[0,1]$ is an arbitrary parameter, and $\kappa_i$ is the
degree of the node $i\in V$. The operator
$f\left(\bf{A}\right)$ is the transition operator of "lazy" random
walks for which
 a random walker stays
in the initial vertex
with probability $1-\beta$,
while it moves
to another node
randomly chosen among the nearest neighbors
with
probability $\beta/\kappa_i$.
In particular,
for  $\beta=1$,
 the operator
$f\left(\bf{A}\right)$ describes the standard
random walks
extensively studied in classical surveys
\cite{Lovasz:1993},\cite{Aldous}.

\subsection{Algebraic geometry of linear automorphisms}
\label{subsec:Representaions}
\noindent

The attractiveness of random walks methods
relies on the fact
that the distribution
of the current node of
any undirected  non-bipartite graph
after $t\gg 1$ steps
tends to a
well-defined stationary distribution,
\begin{equation}
\label{stationary_pi}
\pi_i\,=\,\frac{\,\,\kappa_i\,\,}{\,\,\sum_{j\in V}\kappa_j\,\,},
\end{equation}
which is uniform
if the graph is regular.
The stationary distribution of random
walks (\ref{stationary_pi}) defines a unique measure
on the set of nodes $V$
 with respect to which the
transition operator
((\ref{T_beta}) for $\beta=1$)
is self-adjoint,
\begin{equation}
\label{self_adj}
T\,=\,\frac 12\left( \pi^{1/2}\,f\,\pi^{-1/2}+\pi^{-1/2}\,f^{\top}\,\pi^{1/2}\right),
\end{equation}
where
$f^\top$ is the adjoint operator,
and $\pi$ is defined as
the diagonal matrix
$\pi=\mathrm{diag}\left(\pi_1,\ldots,\pi_N\right)$.

In the canonical basis associated to the  nodes,
$\left\{\left(0,\ldots,1_i,\ldots, 0\right)\right\}_{i=1}^N$,
with $1$ at the $i$-th position, for every node $i\in G$,
the self-adjoint operator (\ref{self_adj})
is represented by a real positive stochastic matrix.
 It follows from
 the
 Perron-Frobenius theorem \cite{Horn:1990} that
its maximal
 eigenvalue is simple and equals $\mu_1=1$,
and
all components of the  eigenvector
    $\psi_1$ belonging to the largest eigenvalue $\mu_1$
are   positive, $\psi_{1,i}=\sqrt{\pi_i}>0$, $\forall i\in V$.
All eigenvalues are
real
 $1=\mu_1>\mu_2\geq\ldots\geq\mu_N\geq -1$,
with orthonormal eigenvectors  $\left\{\psi_k\right\}_{k=1}^N$,
\begin{equation}
\label{normalization}
\sum_{i=1}^N\,\psi^2_{k,i}\,=\,\sum_{k=1}^N\,\psi^2_{k,i}\,=\,1,
\end{equation}
forming an orthonormal basis
in Hilbert space $\mathcal{H}(V)$.
For eigenvalues of  algebraic multiplicity $r>1$, a number of
linearly independent orthonormal eigenvectors can be chosen to span
the associated eigenspace.

Eigenvectors map the nodes of the graph onto the
$(N-1)$-dimensional unit hyper-sphere, $\psi_k:V\to S_1^{(N-1)}$.
Let us denote the matrix in which the eigenvectors $\psi_k$ are
columns by $\bf \Psi$. It is clear that ${\bf \Psi}\in O(N)$ where
$O(N) $ is the group of all orthogonal $N\times N$-matrices  since
 ${\bf \Psi}^{\top}{\bf \Psi}={\bf 1}$ and $\det {\bf \Psi}=\pm 1$.
 The orthogonal matrix ${\bf \Psi}$ has the skew-symmetric property,
$\Psi_{n,m}=-\Psi_{m,n}$, and therefore contains just  $N(N-1)/2$
independent entries which are determined by $N-1$ autonomous
angular coordinates $\theta_{l}$, $l=1,\ldots N-1$. If we denote
the degree of the node $i\in V$ by
$\kappa_i=2|E|\psi^2_{1,i}$, then the angular
coordinates can be defined by
\begin{equation}
\label{tan2}
\tan^2 \theta_{l+1}\,=\,\frac {\,\,1\,\,}{\,\,\kappa^2_{l+1}\,\,}\,\sum_{s=1}^l\, \kappa^2_s,\quad
l=1,\ldots N-2.
\end{equation}
Then the entries of the first eigenvector are given by the
standard relations between the Cartesian and the hyper-spherical
coordinates on the unit hyper-sphere,
\begin{equation}
\label{stationary_point}
\begin{array}{lcl}
\psi_{1,1}&=&\cos \theta_1,\\
&\ldots & \\
\psi_{1,N-1}&=&\cos \theta_{N-1} \prod_{s=1}^{N-2}\sin\theta_s,\\
\psi_{1,N}&=&\prod_{s=1}^{N-1}\sin\theta_s.
\end{array}
\end{equation}
Since all components
of the first eigenvector $\psi_1=\sqrt{\pi}$
 are positive, we can construct
the stereographic projection from the unit hyper-sphere
$S_1^{N-1}$ onto the  hyper-plane $P\mathbb{R}_{\pi}^{(N-1)}$,
with the use of
the point $\psi_1=\left(\psi_{1,1},\ldots\psi_{1,N}\right)$
as the center of projection.
The projective plane $P\mathbb{R}_{\pi}^{(N-1)}$ is then
spanned by the basis vectors,
\begin{equation}
\label{homogen}
\psi'_k\,=\,\left\{\frac{\,\psi_{k,i}\,}{\,\psi_{1,i}\,}\right\}, \quad k\,=\,2,\ldots, N,
 \quad i\in V,
\end{equation}
so that any vector
$\pi^{-1/2}{\bf v}\in P\mathbb{R}_{\pi}^{(N-1)}$
can be expanded into
$\pi^{-1/2}\mathbf{v}=\sum_{k=2}^N\mathbf{v}\mathcal{P}_k$
where
\begin{equation}
\label{project}
\mathcal{P}_k\,\equiv \,\left|\psi'_k\right\rangle\left\langle\psi'_k\right|
\end{equation}
is the projection operator onto the space associated to the
basis vector $\psi'_k$.
The set of all isolated vertices $\mathfrak{P}$ of the graph $G(V,E)$
for which $\pi_{\mathfrak{P}}=0$ play the role of the plane at infinity,
 away from which
we can use the basis defined by (\ref{homogen})
 as the  Cartesian system.

\subsection{Linear automorphisms and probabilities}
\label{subsec:Probabilities}
\noindent

The first eigenvector
    $\psi_1$ belonging to the largest eigenvalue $\mu_1=1$
of (\ref{self_adj}) is nothing else as the Perron-Frobenius
eigenvector which determines the stationary distribution of random
walks \cite{Lovasz:1993},
\begin{equation}
\label{psi_01}
\psi_1
\,\widehat{ T}\, =\,
\psi_1,
\quad \psi_{1,i}^2\,=\,\pi_i,
\end{equation}
and the
Euclidean norm
in the orthogonal complement of $\psi_1$,
$
\sum_{s=2}^N\psi_{s,i}^2\,=\,1-\pi_i,
$
is nothing else as
the probability
that a random walker is not found in $i$.
In a continuous time jump process,
the transition time $\tau$ is a discrete random variable,
 with mean $1$,
 distributed with respect to the Poisson distribution
$\mathrm{Po}(t)$.
The transition
 probability for the continuous time jump process
is
\begin{equation}
\label{limit2}
\begin{array}{lcll}
p^{t}_{ij}&=& \pi_j+\sum_{k=2}^N \psi_{ki}\psi_{kj}\sum_{\tau=0}^\infty
\mu_k^\tau \frac{t^\tau e^{-t}}{\tau!} & \\
 & =& \pi_j+ \sum_{k=2}^N e^{-t\left(1-\mu_k\right)}\,\psi_{ki}\psi_{kj},& i,j\in V.
\end{array}
\end{equation}
Therefore, eigenvectors from the orthogonal complement of $\psi_1$
determine a relaxation process toward the stationary
distribution  of random walks $\pi$
described by the set of characteristic time scale
of the transient process,
\begin{equation}
\label{decay_time}
\tau_k=\left(1-\mu_k\right) ^{-1}, \quad k\geq 2.
\end{equation}
For each
$n=0,\ldots, N$, we can construct a new vector space
$\bigwedge ^n
S_1^{N-1}$, where the symbol $\wedge$
denotes the standard wedge product of vectors in $S_1^{N-1}$.
It is clear that  $\bigwedge ^0
S_1^{N-1} = \mathbb{R}$, $\bigwedge ^1 S_1^{N-1} =
S_1^{N-1} $, and $\bigwedge ^k S_1^{N-1} $ ($2\leq k\leq N$)
consists of all sums $\sum
a\,\alpha_1\wedge\alpha_2\wedge\ldots\wedge\alpha^k$ where $a\in
\mathbb{R}, $ $\alpha_i\in S_1^{N-1}$.

Given an ordered set of
 indices $\mathcal{I}=\left\{k_1,k_2,\ldots,k_n\right\}$,
$1\leq k_1<k_2<\ldots<k_n\leq N$,
the forms
\begin{equation}
\label{wedge_basis}
\psi_{\mathcal{I}}\,=\, \psi_{k_1}\wedge\ldots\wedge \psi_{k_n}
\end{equation}
 over
all possible
$\left(
\begin{array}{c}
N \\
n
\end{array}
\right)$
ordered sets $\mathcal{I}$
is an orthonormal basis in $\bigwedge ^n \mathbb{R}^N$.
As usual, the determinant (\ref{wedge_basis}) can
be considered as a linear and signed version of hyper-volume:
the absolute value of the determinant of real vectors
 is equal to the volume of the parallelepiped spanned by those vectors.
It is clear that $\left|\det {\bf \Psi}\right|=1$, and moreover
the natural  normalization condition holds,
\begin{equation}
\label{normalization2}
\sum_{\{\mathcal{I}\}}\psi_{\mathcal{I}}^2\,=\,1,
\end{equation}
in which the sum is taken over all possible
$\left(
\begin{array}{c}
N \\
n
\end{array}
\right)$
ordered sets of indices $\mathcal{I}$.
The squared hyper-volumes $\psi_{\mathcal{I}}^2$ can
therefore be interpreted as probabilities
to find a random walker
contributing to the
relaxation processes
determined by the eigenmodes
$\mathcal{I}=\left\{ k_1,k_2,\ldots,k_n\right\}$,
\begin{equation}
\label{probab}
\Pr \left[k_1,k_2,\ldots,k_n\right]\,=\,\psi_{\mathcal{I}}^2.
\end{equation}
The obvious and simplest example is the
 probability to observe a random walker in $v$ during infinite time,
$\left.\Pr \left[1\right]\right|_{v\in V}=\psi_{1,v}^2=\pi_v,$ $\sum_v\pi_v=1,$
  given by the stationary distribution of random walks
over nodes.

\subsection{Linear automorphisms and first hitting time}
\label{subsec:Metric}
\noindent

In  graph theory, the distance between two vertices in a
connected graph is the number of edges in a shortest path connecting them.
It is also known as the geodesic distance
  because it is the length of the graph geodesic between those two vertices.
However, given a random walk defined on the undirected connected
 graph $G(V,E)$, each path in that can be
characterized by a certain probability to be followed by random walkers,
so that the shortest path between two nodes
may be different from the most probable one.
In fact, while travelling from the vertex $i\in V$ to the vertex
$j\in V$, random walkers follow with different probabilities
all possible paths that connect them. Hence,
 while discussing linear automorphisms of undirected connected graphs,
it seems natural to generalize the notion of distance
 between two vertices in the
 graph in a probabilistic sense,
namely as the expected number of random steps required to random walkers
starting at $i\in V$
in order to reach $j\in V$ for the first time,
\begin{equation}
\label{hitting}
D(i,j)\,=\,t_{ij}.
\end{equation}
  The above defined quantity is called the first hitting time, \cite{Lovasz:1993}.

The standard computation of the first hitting time
which can be found in \cite{Lovasz:1993}
is based on the fact that while travelling  from $i\in V$ to $j\in V$,
the first step takes us to a neighbor
 $v\in V$ of $i$ and then we have to reach
$j$ from there,
\begin{equation}
\label{hitting_fact}
t_{ij}\,=\,1+\frac{1}{\,\,\kappa_i\,\,}\,\sum_{i\sim v}\,t_{vj},
\end{equation}
together with the obvious requirement $t_{ii}=0.$

Alternatively, we note that random walks
defined on finite undirected graphs
can be naturally related to a diffusion process
which describes the
dynamics of a large number of random walkers.
The symmetric diffusion process corresponding
to the self-adjoint transition operator of
random walks (\ref{self_adj})
 determines the time evolution of the normalized expected number
of random walkers, $\pi^{-1/2}{\bf n}(t)\in V\times \mathbb{N}$,
\begin{equation}
\label{diffusion}
\dot{\bf n}\,=\, {  L}{\bf n},\quad  {  L}\,\equiv\, {  1}- {  T}
\end{equation}
where $ {L}$ is the normalized Laplace operator defined on $G$.
Eigenvalues of $ {L}$ and $T$ are
simply related by
$\lambda_k=1-\mu_k$, $k=1,\ldots, N$, while
 eigenvectors of both operators are identical.
The analysis of spectral properties of the operator
(\ref{diffusion}) is widely used in the spectral graph
theory, \cite{Chung:1997,Blanchard:2008}.

The expected first hitting time
 can formally be calculated as
\begin{equation}
\label{formal_solution}
t_{ij}\,=\,\int_{t_i}^{t_j}\,\delta t\,=
\,\int\,\frac{\,\,\delta n_{ij}\,\,}{\,\,{L_{ij} n_j}\,\,}
\,=\,\int\,\frac {\mathbb{P}_{ij}}{\,\,L_{ij}\,\,}, \quad \mathbb{P}_{ij}\,\equiv\frac{\,\, n_i\,\delta n_{ij}\,\,}{\,\,{\|{\bf n}\|^2}\,\,},
\end{equation}
if the inverse
Laplace operator $1/L\equiv  \sum_{n\geq 1} {T}^n $ exists.
The r.h.s. of (\ref{formal_solution}) is nothing else but a convolution
of the projection operator onto the difference direction vector
 $\mathbb{P}_{ij}$ with the
Fredholm kernel of the Laplace equation.

\begin{figure}[ht]
\label{Fig1}
 \noindent
\epsfig{file=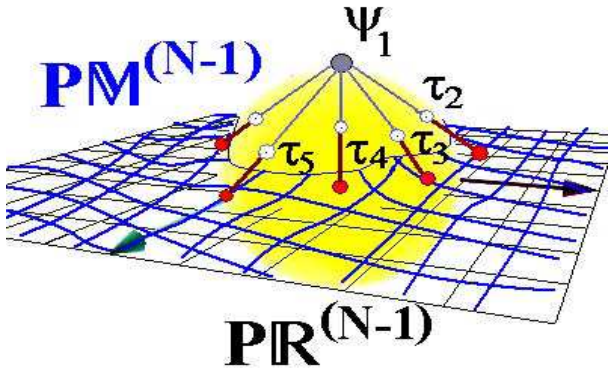, width=8cm, height =7cm}
\caption{Schematic representation of a projective action of the inverse Laplace operator,
${L}^{-1}:S_1^{N-1}-\left\{\psi_1\right\}\to P\mathbb{M}^{N-1}$.}
\end{figure}

Although the Laplace operator $L$
has a trivial eigenvalue
$\lambda_1=1-\mu_1=0$ and therefore is
not invertible over $S_1^{N-1}$,
 it is invertible
over  $S_1^{N-1}-\left\{\psi_1\right\}$, the orthogonal complement
of the first eigenvector $\psi_1$ (belonging to the largest
eigenvalue of (\ref{self_adj}) $\mu_1=1$) that corresponds to the
transient process of random walks toward the stationary
distribution $\pi$. The orthogonal complement is homeomorphic to
the projective hyper-plane $ P\mathbb{R}_{\pi} ^{(N-1)}$
constructed by linearly mapping points of the unit hyper-sphere
$S_1^{N-1}$ from $\psi_1$ as the center of projection. The inverse
Laplace operator defined by the kernel
 \begin{equation}
\label{Green}
 {L}^{-1}\,=\,
\sum_{k=2}^N\frac{\,\,\mathcal{P}_k\,\,}{\,\,1-\mu_k\,\,},
\end{equation}
in which $\mathcal{P}_k$ is the projector (\ref{project}),
is itself a projection operator
${L}^{-1}:S_1^{N-1}-\left\{\psi_1\right\}\to P\mathbb{M}^{N-1}$
where the $k^{\mathrm{th}}$ coordinate
of the projective manifold $P\mathbb{M}^{N-1}$
is subjected to the dilatation
$\tau_k=\left(1-\mu_k\right)^{-1}$, $k=2,\ldots, N$ (see Fig.~\ref{Fig1}).
The kernel (\ref{Green}) defines
 the Green function (or the Fredholm kernel)
describing long-range interactions between
eigenmodes of the diffusion process
induced by the
graph structure.

The convolution  with  Green's function gives solutions to
inhomogeneous Laplace equations.
It is remarkable that such a convolution with
the projection operator $\mathbb{P}_{ij}$  gives for the
first hitting time the expression
\begin{equation}
\label{formal_solution2}
t_{ij}\,=\, \sum_{k=2}^N \,\frac{\,\,\psi'_{ki}{}^2-\psi'_{ki}\psi'_{kj}\,\,}{\,\,1-\mu_k\,\,},
\end{equation}
which coincides with the result of \cite{Lovasz:1993}.

It is well known \cite{Aldous,Lovasz:1993} that the matrix of first hitting times
$t_{ij}$ is not symmetric,
 even  for a regular graph, and therefore in general $D(i,j)\ne D(j,i)$.
However, one can consider its symmetrized analog,
\begin{equation}
\label{commute}
d(i,j)\,=\,D(i,j)+D(j,i)\,=\,\sum_{k=2}^N \,\frac{\,\,\psi'_{ki}{}^2- \psi'_{kj}{}^2\,\,}{\,\,1-\mu_k\,\,},
\end{equation}
known as commute time \cite{Lovasz:1993}, the
 expected number of steps required for a random
walker starting at $i$ to visit $j$ for the first time and then to
return back to $i$ for the first time. It is easy to check that
commute time defined by (\ref{commute}) satisfies all distance
axioms.

\subsection{The Gram matrix of graph nodes}
\label{subsec:Manifold}
\noindent

Each vertex  $i\in V$ of the graph $G$
 has an image in $P\mathbb{M}^{N-1}$
determined by the vector
\begin{equation}
\label{image}
{\bf v}_i\,=\,
\left(
\frac{\,\,\psi'_{2,i}\,\,}{\,\,\sqrt{\left(1-\mu_2\right)}\,\,},\ldots,
\frac{\,\,\psi'_{N,i}\,\,}{\,\,\sqrt{\left(1-\mu_N\right)}\,\,}
\right).
\end{equation}
The Gram matrix defined on the set of all vectors $\left\{{\bf v}_i\right\}$, $i\in V$,
is given by
\begin{equation}
\label{Gram}
\mathcal{G}_{ij}\,=\,\sum_{k=2}^N\,
\frac{\,\,\psi'_{k,i}\psi'_{k,j}\,\,}{\,\,\left(1-\mu_k\right)\,\,}.
\end{equation}
The diagonal elements  $\mathcal{G}_{ii}$ are the first-passage times \cite{Lovasz:1993},
the
expected numbers of steps required for  random walkers to reach the node
 $i$ for the first time  starting from any node $l\in V$
 randomly chosen among all
  nodes of the graph $G$ with probability $\pi_l$.
The elements $\mathcal{G}_{ij}$, $i\ne j$,
estimate
the
expected overlap of random walks
towards the nodes $i$ and $j$ starting from a
node $l\in V$
randomly chosen among all
  nodes of the graph $G$ with probability $\pi_l$.

Discovering   important nodes
and quantifying differences between them
 in a graph
is not easy, since
the graph, in general,
does not possess
the structure of Euclidean space.
The first passage time,  $\mathcal{G}_{ii}$,
 can be directly used in order to characterize
the level of accessibility of the node $i$ in the graph $G$.
Various properties of first-passage times
have been recently studied in
 concern with the  {traffic flow
forecasting} \cite{Sun:2005},  in order
  to model the wireless
terminal movements in a  {cellular wireless
  network} \cite{Jabbari:1999},
in a statistical test for the presence
   of a random walk component
in the  {repeat sales price
   models} in house prices \cite{Hill:1999},
in the growth modelling
   of urban agglomerations \cite{Pica:2006},
and in many other
   works where random walks
have been considered directly on
    city plans and physical landscapes.

In contrast to all previous studies, in our approach, we use
first-passage times of discrete time random
 walks in order to investigate
the configuration of urban places represented by means of the
spatial graph.

\section{Linear automorphisms of urban environments. A case of study}
\label{sec:urban}
\noindent

\subsection{Spectra of cities}
\label{subsec:spectra}
\noindent

If we take many, many random numbers from an interval of all real
numbers symmetric with respect to a unit and calculate the sample
mean in each case, then the distribution of these sample means
will be approximately normal in shape and centered at 1 provided
the size of samples was large. The probability density function of
a normal distribution forms a symmetrical bell-shaped curve
highest at the  mean value indicating that in a random selection
of the numbers around the mean (1) have a higher probability of
being selected than those far away from the mean. Maximizing
information entropy among all distributions with known mean and
variance, the normal distribution arises in many areas of
statistics.

\begin{figure}[ht]
 \noindent
\begin{center}
\epsfig{file=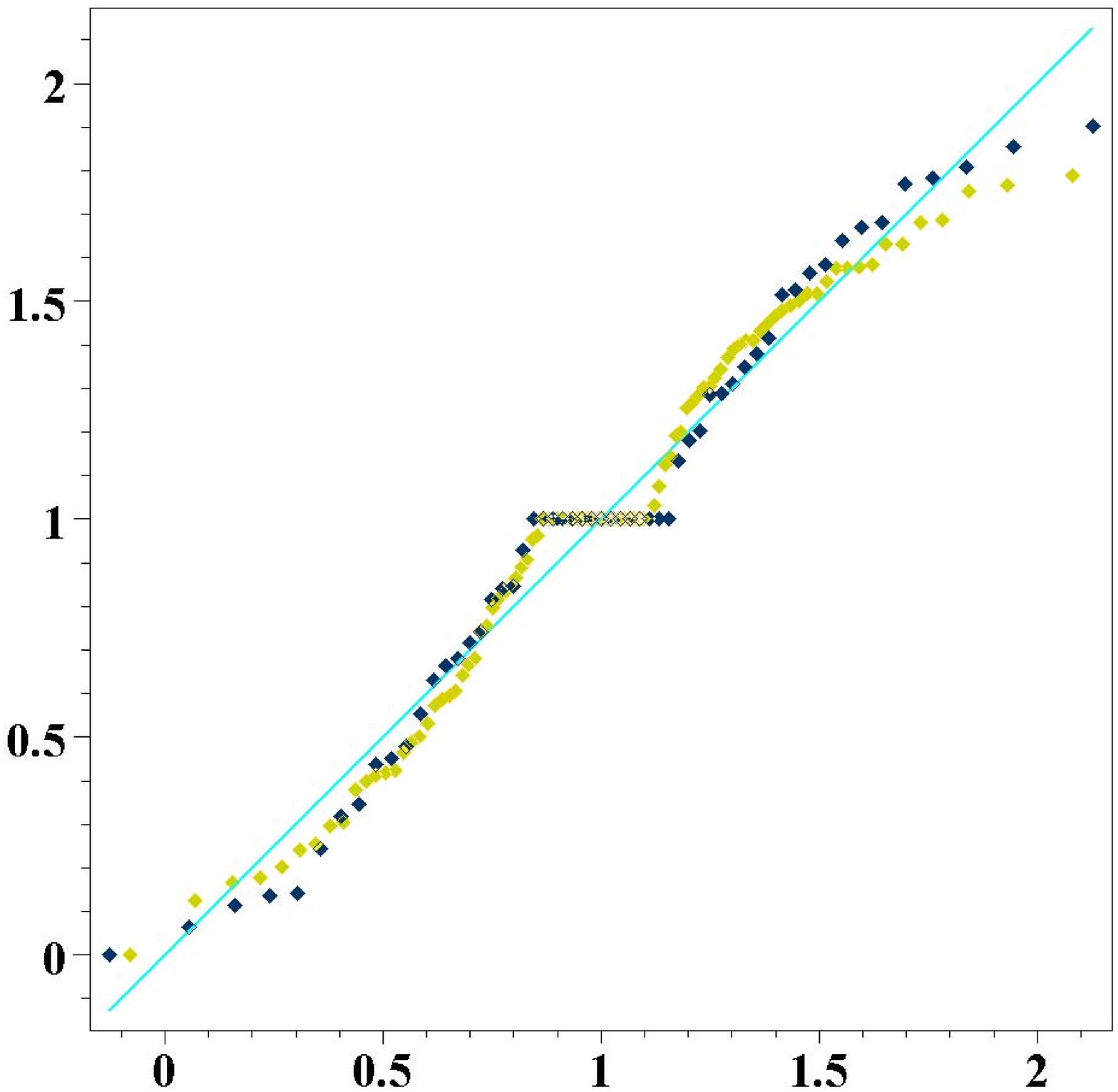,  angle= 0,width =9cm, height =8cm}
  \end{center}
\caption{\small The probability-probability plot of the
normal distribution (on the horizontal axis) against
the empirical distribution of eigenvalues
in the spectra of the city canal networks in Venice and Amsterdam.
The diagonal line $y = x$ is set for a reference.}
 \label{Fig2}
\end{figure}

It is interesting to compare the empirical
distributions of eigenvalues
of the normalized Laplace operator (\ref{diffusion})
defined on the spatial graphs of compact urban patterns -- the
spectra of  cities --
 with the
normal distribution centered at 1. In Fig.~\ref{Fig2}, we have
shown a probability-probability plot of the normal distribution
(on the horizontal axis) against the  empirical distribution of
eigenvalues in the city spectra (the  normal plot)
of the city canal networks in Venice (96 canals) and Amsterdam (57 canals).
A random sample of the normal
distribution, having size equal to the number of eigenvalues in
the spectrum has been be generated, sorted ascendingly, and
plotted against
 the  response of
the empirical distribution of city eigenvalues.
The  spectra of canal
maintained
in the compact urban patterns of Venice and Amsterdam
look also amazingly alike
and are obviously tied to the normal distribution,
although these canals had
  been founded in
in the dissimilar geographical regions
 and for the different purposes.
While the Venetian canals
mostly serve the function of
transportation routs
 between the distinct districts
of the gradually growing naval capital of the Mediterranean
region, the concentric web
 of Amsterdam gratchen
 had been built in order to defend the city.

It is remarkable that
the spectral density distributions
shown in Fig.~\ref{Fig2}
are dramatically dissimilar to those
reported
 for the random
graphs of Erd\"{o}s and R\'{e}nyi
studied by \cite{Farkas:2001,Farkas:2002}.
The classical
 {Wigner semicircle distribution}
arises as the limiting
 distribution of eigenvalues
of many random symmetric matrices as the size of the
 matrix approaches infinity, \cite{Sinai:1998}.
 The eigenvalues of the
normalized Laplace operator in a random scale-free graph also
 follow the semicircle law \cite{Chung:2003}.
City spectra reveal the profound structural
dissimilarity between urban networks and networks of other types
studied before.

\subsection{First-passage times to ghettos}
\label{subsec:FPT}
\noindent

The phenomenon of clustering of minorities, especially that of newly
 arrived immigrants, is well documented \cite{Wirth} (the reference appears
 in \cite{Vaughan01}).
Clustering is considering to be beneficial for mutual support and
for the sustenance of cultural and religious activities. At the
same time, clustering and the subsequent physical segregation of
minority groups would cause their economic marginalization.
The spatial analysis of the immigrant
quarters \cite{Vaughan01}  and the
study of London's changes over 100 years \cite{Vaughan02}
shows that they were significantly more
segregated from the neighboring areas, in particular, the number
of street turning away from the quarters to the city centers were
found to be less than in the other inner-city areas being
usually socially barricaded by railways, canals and industries.
It has been suggested \cite{Language} that space structure and its
impact on  movement are critical to the link between the built
environment and its social functioning. Spatial structures creating a local situation
in which there is no relation between movements inside the spatial
pattern and outside it and the lack of natural space occupancy
become associated with the social misuse of the structurally
abandoned spaces.

\begin{figure}[ht]
 \noindent
\begin{center}
\epsfig{file=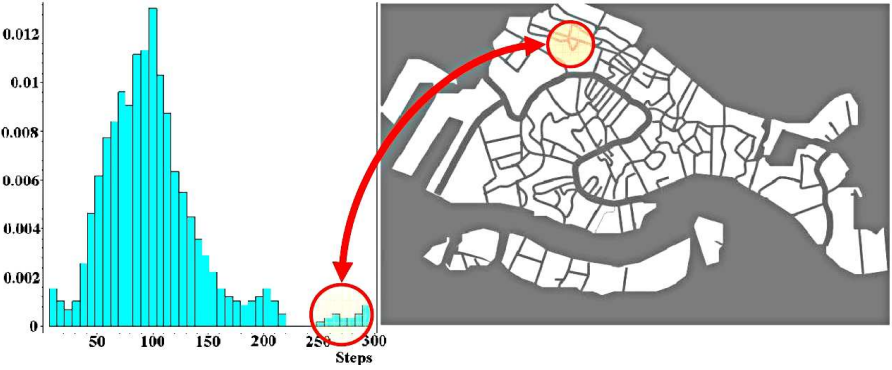,  angle= 0,width =12cm, height =5cm}
  \end{center}
\caption{\small  The Venetian Ghetto jumped out
as by far the most isolated, despite being apparently well connected to the rest of the city. }
 \label{Fig3}
\end{figure}

We have analyzed the first-passage times to individual canals in the
spatial graph of the canal network in Venice.
The distribution of numbers of canals
over the range of the first--passage time values is represented
by a histogram shown in Fig.~\ref{Fig3}.left.
 The height of each bar in the histogram
is proportional to the number of canals in the
 canal network of Venice for which the first--passage
times fall into the disjoint intervals (known as bins).
Not surprisingly,
the Grand Canal, the giant Giudecca Canal
 and the Venetian lagoon are the most connected.
In contrast,  the Venetian Ghetto (see Fig.~\ref{Fig3}.right) -- jumped out
as by far the most isolated, despite being apparently well connected to the rest of the city --
 on average, it took 300 random steps to reach, far more than the average of 100 steps for other places in Venice.

The Ghetto was created in March 1516 to separate Jews from the Christian majority of Venice. It persisted until
1797, when Napoleon conquered the city and demolished the Ghetto's gates.
Now it is abandoned.

\subsection{Random walks estimate land value in Manhattan}
\label{subsec:manhattan}
\noindent

The notion of
isolation
 acquires
the statistical interpretation by means of random walks. The
first-passage times in the city vary strongly from  location to
location. Those places characterized by the shortest first-passage
times are easy to reach while very many random steps would be
required in order to get into a statistically isolated site.

Being a global characteristic
of a node in the graph,
the first-passage time
assigns  absolute scores
to all nodes
 based on the probability
 of paths they provide
 for random walkers.
The first-passage time
can therefore be considered
as a natural
statistical
centrality measure of
 the vertex within the graph, \cite{Blanchard:2008}.

 \begin{figure}[ht]
 \noindent
\begin{center}
\epsfig{file=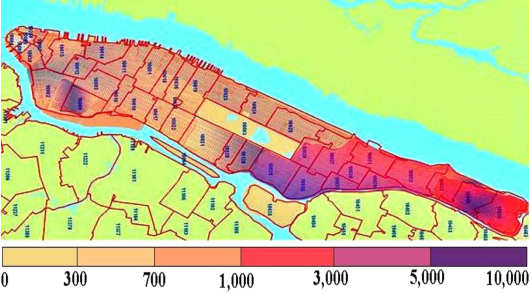, angle= 0,width =7cm, height =4cm}
  \end{center}
\caption{\small Isolation map of Manhattan. Isolation is measured
by  first-passage times to the places.  Darker color corresponds
to longer first-passage times.}
 \label{Fig2_Isolation}
\end{figure}

A  visual pattern
displayed on Fig.~\ref{Fig2_Isolation}
represents the pattern of structural
  isolation (quantified by the first-passage times)
 in Manhattan (darker color corresponds to longer first-passage times).
It is interesting to note that the  {spatial distribution of
isolation} in the urban pattern of Manhattan
(Fig.~\ref{Fig2_Isolation})
 shows a qualitative agreement with the map
of the  tax assessment  value of the land in Manhattan reported by
B. Rankin (2006) in the framework of the RADICAL CARTOGRAPHY
project being practically a negative image of that.

Recently, we have discussed in \cite{Blanchard:2008}
that distributions of
 various social variables
(such as the mean household income and prison expenditures in
different zip code areas) may demonstrate the striking spatial
patterns which can be analyzed by means of random walks. In the
present work, we analyze the spatial distribution of the tax
assessment rate (TAR) in Manhattan.

The assessment tax relies upon a special enhancement made up of
the land or site value and differs from the market value
estimating a relative wealth of the place  within the city
commonly refereed to as the 'unearned' increment of land use,
\cite{Bolton:1922}. The rate of appreciation in value of land is
affected by a variety of conditions, for example it may depend
upon other property in the same locality, will be due to a
legitimate demand for a site, and for occupancy and height of a
building upon it.

 \begin{figure}[ht]
 \noindent
\begin{center}
\epsfig{file=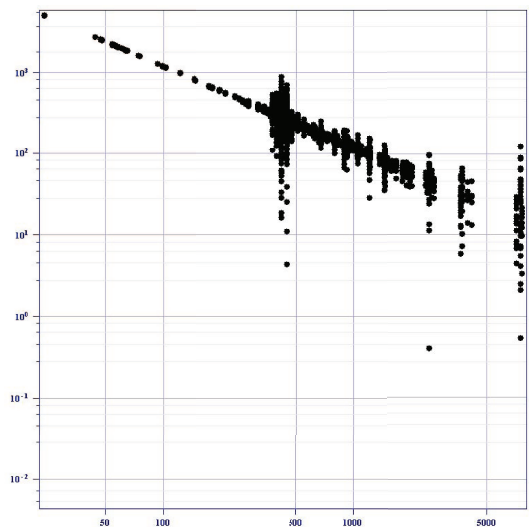, angle= 0,width =7cm, height =7cm}
  \end{center}
\caption{\small Tax assessment rate (TAR) of places in Manhattan
 (\$/fit${}^{2}$) is shown in the logarithmic scale
vs.
the first--passage times (FPT) to them.  }
 \label{Fig2_prices}
\end{figure}

The current tax assessment system enacted in 1981
in the city of New York
 classifies all real estate parcels into four classes subjected
to the different tax rates set by the legislature:
(i) primarily residential condominiums; (ii) other residential property;
(iii) real estate of utility corporations and special franchise properties;
(iv) all other properties, such as stores, warehouses, hotels, etc.
However, the scarcity
of physical space in the compact urban pattern on the island of Manhattan
will naturally set some increase of value on all desirably located
land as being a restricted commodity.
Furthermore, regulatory constrains on housing supply exerted on housing prices
by the state and the
city in the form of 'zoning taxes'
are responsible for
converting the property tax system in a complicated mess
of interlocking influences and for much of the high cost of housing
in Manhattan, \cite{Glaeser:2003}.

Being intrigued with the
 likeness of
the tax
assessment map and the map of isolation in Manhattan,
 we have mapped the TAR figures publicly available
through the Office of the Surveyor at
the Manhattan Business Center onto the
 data on first-passage times to the corresponding
 places.
The resulting plot is shown in Fig.~\ref{Fig2_prices}, in the
logarithmic scale. The data presented in Fig.~\ref{Fig2_prices}
positively relates the geographic accessibility of places in
Manhattan
 with their 'unearned increments'
estimated by means of the increasing burden of taxation.
The inverse linear pattern dominating the data
is best fitted by the simple hyperbolic relation between
the tax assessment rate (TAR)
and the value of first--passage time (FPT),
\begin{equation}
\mathrm{TAR}\,\propto\,\frac{c}{\,\,\mathrm{FPT}\,\,},
\end{equation}
in which $c\simeq 120,000\,\, \$\times\mathrm{Step}/\mathrm{fit}^2$ is a fitting
constant.

\ack{This work was supported by the Volkswagen Foundation (Germany)
in the framework of the project "Network formation rules, random
set graphs and generalized epidemic processes" (Contract no Az.:
I/82 418).}

\end{document}